\documentstyle[12pt]{article}
\newcommand{\bea}{\begin{eqnarray}}
\newcommand{\be}{\begin{equation}}
\newcommand{\eea}{\end{eqnarray}}
\newcommand{\ee}{\end{equation}}
\def\nn{\nonumber}

\def\a{\alpha}
\def\b{\beta}

\def\d{\delta}
\def\e{\epsilon}

\def\l{\lambda}

\def\L{\Lambda}

\def\cd{{\cal D}}

\def\cg{{\cal G}}
\def\ch{{\cal H}}

\def\co{{\cal O}}

\begin{document}
\pagestyle{empty} \large \noindent . \vspace*{5mm} \\
CGPG-94/1-3\\January 1994\\
\begin{center} \LARGE \bf \vspace*{35mm}
Towards a unification of gravity and Yang-Mills theory\\
\vspace*{20mm} \large \bf
Subenoy Chakraborty\dag \ddag  and Peter Peld\'{a}n\dag
\footnote{Email address: peldan@phys.psu.edu}\\
\vspace*{5mm} \large
\dag Center for Gravitational Physics and Geometry\\
The Pennsylvania State University, University Park, PA 16802, USA\\
\vspace*{5mm}\ddag Department of Mathematics\\
Jadavpur University, Calcutta-700032, India\\
\vspace*{20mm} \Large \bf
Abstract\\
\end{center} \normalsize
We introduce a gauge and diffeomorphism invariant theory on Yang-Mills phase
space. The theory is well defined for an arbitrary gauge group with an
invariant bilinear form, it contains only first class constraints, and the
spacetime metric has a simple form in terms of the phase space variables. With
gauge group $SO(3,C)$, the theory equals the Ashtekar formulation of gravity
with a cosmological constant. For Lorentzian signature, the theory is
complex, and we have not found any good reality conditions. In the Euclidean
signature case, everything is real. In a weak field expansion around de
Sitter spacetime, the theory is shown to give the conventional Yang-Mills
theory to the lowest order in the fields.

\newpage \pagestyle{plain}
Comparing the two dominating Hamiltonian formulations of (3+1)-dimensional
Einstein gravity; the ADM-formulation \cite{ADM} and the Ashtekar formulation
\cite{Ash1}, it is quite clear that gravity in (3+1)-dimensions (and
(2+1)-dimensions) seems to prefer the Yang-Mills phase-space compared to the
geometrodynamical phase-space. This is mainly due to the fact that the
constraints in the theory simplify a lot in the transition to the Ashtekar
formulation. There is, however, an ugly spot in this otherwise very beautiful
formulation: the need for complex fields and complicated reality conditions.
The Ashtekar Hamiltonian for Einstein gravity is known to exist for (2+1)-
and (3+1)-dimensional Lorentzian and Euclidean gravity, but it is only for
the (3+1)-dimensional Lorentzian case (the theory we are most interested in)
that we need complex fields!

Another "disadvantage" of the Ashtekar variables can be found in the matter
couplings: although both scalar fields and spinor fields can be beautifully
incorporated into the theory, the coupling to the spin-1 Yang-Mills fields
destroys the simplicity of the constraints. See \cite{ART}. Either the
coupling is non-polynomial or one has to rescale the pure
gravity part of the Hamiltonian constraint by multiplying it with the
determinant of the metric. Both choices will probably
severly complicate the canonical quantization of this theory.\\ \\
These three facts taken together -- (1) gravity prefers the Yang-Mills
phase-space, (2) the theory is complex, and (3) the coupling to spin-1
Yang-Mills fields does not seem natural -- could indicate that there exists
another
underlying more beautiful theory. This would presumably be a
real (non-complex) unified theory of gravity and Yang-Mills theory, for some
gauge
group $G$, such that,
when the larger symmetry is broken down to $G\sim SO(3)\times G^{YM}$, the need
for complex fields appears.\\ \\
In this letter, we will describe a candidate theory for this unification. We
have, however, not been able to find a real theory for the case of Lorentzian
signature of the metric. Otherwise, the theory fulfills the requirements put
on it so far: it is a diffeomorphism and gauge invariant theory valid for any
gauge group which has a non-degenerate invariant bilinear form. Furthermore,
 our model
reduces to conventional Einstein gravity with a cosmological constant if one
uses the gauge group $SO(3,C)$, and in an expansion for weak fields around
de-Sitter spacetime, the theory agrees with conventional Yang-Mills
theory to lowest order.\\ \\

Here is the Hamiltonian for the unified theory\footnote{Compared to
\cite{thesis} and
\cite{arbgg}, the fields are rescaled as follows: $E^{ai}\rightarrow -iE^{ai}$,
$A_{ai}\rightarrow iA_{ai}$, $F^i_{ab}\rightarrow iF^i_{ab}$}:

\bea \ch_{tot}&=& N \ch + N^a \ch_a + \L _i \cg ^i \label{1} \\
\ch&:=&\frac{1}{4}
\e_{abc}\e_{ijk}({\scriptstyle E^{dl}})E^{ai}E^{bj}(B^{ck}+\frac{2i\l}{3}
E^{ck})\approx 0 \\
\ch _a&:=&\frac{1}{2}\e_{abc}E^{bi}B^c_i\approx 0 \\
\cg _i&:=&\cd _a E^a _i=\partial _a E^a _i + f_{ijk}A_a^j E^{ak}\approx 0
\eea
The index-conventions are: $a, b, c, .....$ are spatial indices on the three
dimensional hypersurface, and $i, j, k, .......$ are gauge-indices in the
vector representation, and therefore take the values $1,2, ....N$ where $N$
is the dimension of the Lie-algebra. Gauge-indices are raised and lowered
with an invariant bilinear form of the Lie-algebra (the "group-metric"),
often conveniently chosen
to be the Cartan-Killing form. The basic conjugate fields are $A_{ai}$ and
$E^{bj}$, which satisfy the fundamental Poisson bracket: $\{
A_{ai}(x),E^{bj}(y)\}=\d ^b _a \d ^j _i \d ^3 (x-y)$. $A_{ai}$ is a gauge
connection, and $E^{ai}$ is often referred to as the "electric field". The
other
fields in the theory, $N, N^a, \L _i$ are Lagrange multiplier fields whose
variations impose the constraints $\ch, \ch _a$ and $\cg_i$. $B^{ai}$ is the
"magnetic field": $B^{ai}:=\e^{abc}F^i_{bc}=\e^{abc}(2 \partial _b A^i_c +
f^i{}_{jk}A_b^jA_c^k)$, and $\l$ is the cosmological constant.

The most important ingredient in this formulation
 is $\e_{ijk}({\scriptstyle E^{dl}})$, which is
defined as follows:

\be \e_{ijk}({\scriptstyle
E^{dl}}):=\frac{\e_{abc}E^{a}_iE^b_jE^c_k}{\sqrt{det(E^{dl}E^e_l)}} \label{2}
\ee
It is a generalization of the three dimensional Levi-Civita symbol to higher
dimensional Lie-algebras. If the gauge-group is chosen to be three
dimensional, $\e_{ijk}({\scriptstyle E^{dl}})$ just reduces to the normal
Levi-Civita symbol, and the Hamiltonian (\ref{1}) equals the Ashtekar
Hamiltonian for gravity with a cosmological constant. Note that this $\e
_{ijk}$ needs
a non-zero $det(E^{ai}E^b_i)$ for its definition, or, using the result
(\ref{3})
below; $\e _{ijk}$ is only well defined for non-degenerate metrics.

Now, before one can say that one has a well defined and consistent theory
described by the Hamiltonian (\ref{1}), one must check if the time evolution
of the constraints vanishes weakly (in Dirac's terminology). And, when the
total Hamiltonian is a linear combination of constraints, this corresponds to
checking the constraint algebra. Since $\cg _i$ is the well known Gauss law
constraint, which is known to generate gauge transformations, and $\ch _a$ is
the vector constraint, which is known to generate spatial diffeomorphisms
(modulo gauge transformations), it is an easy task to calculate all the
Poisson
brackets containing these two constraints. See e.g \cite{thesis} for details.
As long as all the constraints are gauge and diffeomorphism covariant,
the Poisson brackets containing $\cg _i$ and $\ch _a$ will all weakly vanish.
This is the case here. The only Poisson bracket left to calculate is thus $\{
\ch [N],\ch [M]\}$. This calculation is a bit messy but it simplifies if one
notes
that the result must be antisymmetric in $N$ and $M$ meaning that only the
terms containing derivatives on these fields survive. Here is the result:

\be \{ \ch [N],\ch [M]\}=\ch _a[E^{ai}E^b_i(M\partial _bN-N\partial _bM)]
\label{3} \ee
where the square brackets denote smearing over the hypersurface:\\ $\ch
[N]:=\int _{\Sigma} d^3x \; N(x)\ch (x)$. Thus, the constraint algebra
closes, and the theory is complete and consistent, in this sense.

Now, according to Hojman et al. \cite{HKT}, the above Poisson bracket can be
used to
read off the spatial metric in the theory. It was shown in ref. \cite{HKT}
that in any canonical formulation of a diffeomorphism invariant theory, with
a metric, the spatial metric on the hypersurface always appears as a structure
function in the Poisson bracket above. This result was derived from
a consistency
requirement on the theory, so although, at this stage,we do not know the
physical importance of this metric, it is the only consistent
choice for a metric. For a more detailed discussion regarding this question,
see \cite{thesis}. Moreover, in ref. \cite{HKT} it was also shown that the
Poisson bracket between the spatial metric and the Hamiltonian constraint
always gives the extrinsic curvature. And, with both the spatial metric and
the extrinsic curvature at hand, it is possible to reconstruct the entire
spacetime metric:

\be \tilde{g}^{\a \b}=\sqrt{-g}g^{\a \b}=\left(\begin{array}{cc}
-\frac{1}{N}& \frac{N^a}{N}\\
\frac{N^a}{N}& -NE^{ai}E^b _i -\frac{N^aN^b}{N}
\end{array} \right) \label{3b} \ee

The fact that the spatial metric has such a simple form in terms of the phase
space variables makes it rather easy to impose the Lorentzian signature
condition on the theory. In ref. \cite{arbgg} another gauge group
generalization of the Ashtekar Hamiltonian was given, but for that model, the
metric had a very complicated dependence on the phase space variables, making
it almost impossible to restrict the metric signature by imposing conditions
on the basic fields. Here, Lorentzian signature corresponds to requiring
$E^{ai}E^b_i$ to be negative definite.

It is at this stage one may start looking for a real theory. The need for
complex fields for the pure gravity case can be found from the Lorentzian
signature condition; for pure gravity, the gauge group is $SO(3,C)$ and the
"group metric" is the positive definite $\d _{ij}$, which means that
$E^{ai}$ must become complex in order for $E^{ai}E^b_i$ to be negative
definite. (If one from the beginning chooses the "group metric" to be $-\d
_{ij}$, the right hand side of (\ref{3}) will change sign, and the
implications of the signature condition are unaltered.) Now, by choosing a
gauge group with an indefinite "group metric", such as e.g $SO(1,3)$, one
would naively believe that the signature condition could be taken care of
without ever having to introduce complex fields. This, however, does not
work. With a negative definite $E^{ai}E^b_i$, the generalized epsilon
(\ref{2}) becomes complex, so although the fields are taken to be real, the
complexification of the theory is introduced by this epsilon. Then, one could
try to rescale the Hamiltonian constraint $\ch$ by multiplying with the
denominator in (\ref{2}). The result is that the right hand side of (\ref{3})
changes so that the densitized spatial metric becomes $-det(E^{ck}E^d_k)
E^{ai}E^b_i$; a metric that by construction is negative definite (for real
fields) and hence corresponds to Euclidean signature.

Thus, the conclusion
from this failure must be that it is \underline{not} possible to
"uncomplexify" the
theory by simply generalizing it to other gauge groups and then pick a group
with an indefinite "group metric". However, this does not mean that the hope
for finding a real unified theory is dead. It may be that the real theory
only exist for a very special gauge group with some "fancy" feature. After
all, that is what we really wants: the theory telling us what gauge group to
choose. So far this is only speculations, what we can say at this stage
 is that
the theory presented in this letter needs complex fields in the Lorentzian
case. The Euclidean case is perfectly all right with real fields. It is given
by (\ref{1}) with the "i" removed from the Hamiltonian constraint.  \\ \\

At this point, we know that the theory described by (\ref{1}) is gauge
invariant
(since the generator of gauge transformations $\cg _i$ is a first class
constraint), we know that the theory is diffeomorphism invariant (since it
has a constraint algebra required for such a theory \cite{HKT}) and we know
that the densitized spacetime metric is given by (\ref{3b}). The question is
then; what kind of physical interpretation can be given to this theory? We
will now show that if we expand this theory to first order around de Sitter
spacetime, the Hamiltonian will coincide with the normal Yang-Mills
Hamiltonian.

 The idea is to, in
(\ref{1}), use a gauge group which is a direct product of $SO(3,C)$ and an
arbitrary Yang-Mills gauge group: $G^{tot}=SO(3,C)\times G^{YM}$. Then, we
will expand the total Hamiltonian for the unified theory (\ref{1}) around the
de Sitter solution, and keep only the lowest order terms. We will denote the
gravitational $SO(3,C)$ gauge indices by $A,B,C,....$ and the Yang-Mills
$G^{YM}$ gauge indices
by $I,J,K,......$. E.g $E^{ai}E^{b}_i=E^{aA}E^{b}_A + E^{aI}E^b_I$.
To perform this expansion, we do not need to use the explicit de Sitter
solution, it suffices to know that

\be B^{aA}=-\frac{2i \l}{3} E^{aA} \label{5} \ee
for the de Sitter solution, in terms of Ashtekar's variables \cite{unpubl}.
We denote the exact solution by putting a bar on the fields, and the
 perturbations around it with lower case letters:

\bea E^{aA}&=&\bar{E}^{aA}+e^{aA} \label{6} \\
     E^{aI}&=&\bar{E}^{aI}+e^{aI}=e^{aI} \nn \\
     B^{aA}&=&\bar{B}^{aA}+b^{aA}=-\frac{2i \l}{3}\bar{E}^{aA}+b^{aA} \nn \\
     B^{aI}&=&\bar{B}^{aI}+b^{aI}=b^{aI} \nn \\
     N&=&\bar{N}+n \nn \\
     N^a&=&\bar{N}^a + n^a = n^a \nn \\
     \L _i&=&\bar{\L}_i + \l _i \nn \eea
Here, we have used (\ref{5}) and the fact that the Yang-Mils fields vanishes
in the de Sitter solution. We have also picked a coordinate-system where
$\bar{N}^a=0$. Now, weak-field expansion here means: $e^a_A \ll \bar{E}^a_A$,
$b^a _A \ll \l \bar{E}^a_A$, $e^{aI}e^b_I \ll
\bar{q}^{ab}:=\bar{E}^{aA}\bar{E}^b_A$ and $e^{aI}b^b_I \ll \l \bar{E}^{aA}
\bar{E}^b_A$. Since we are mainly interested in the Yang-Mills part, we do
not write out the pure gravity perturbation (although, it is a straightforward
task to use the above expansion to get the gravitational part as well):

\bea N\ch&=&\bar{N}\ch ^{(2)}_{YM}+N\ch _{GR}+ \co (e^4,b^4,...)
\nn \\
N^a\ch _a&=&N^a\ch _a^{GR} + \co (e^4,b^4,....) \label{10} \\
\L ^i\cg _i&=&\l^I\cg ^{(2)YM}_I+\L ^A\cg _A^{GR} + \co
(e^4, b^4,....) \nn \eea
where
\bea \bar{N}\ch ^{(2)}_{YM}&=&\frac{\bar{N}}{4}\frac{1}{\sqrt{\mid
\bar{q}^{ab}\mid}}(\e _{abc}\e _{def}\bar{q}^{ad}\bar{q}^{be}e^{cI}(b^f_I +
\frac{2 i \l}{3} e^f_I)) \label{20} \\
\l^I\cg ^{(2)YM}_I&=&\l ^I \cd _a e^a_I=\l ^I(\partial _a e^a_I +
f_{IJK}a_a^Je^{aK}) \nn \eea
and the $\co (e^4, b^4,....)$ term includes all the higher order terms. $\mid
\bar{q}^{ab}\mid$ is the determinant of the spatial metric $\bar{q}^{ab}$.
 Now,
compare the expressions in (\ref{20}) to the conventional Yang-Mills
total Hamiltonian on any fixed background \cite{ART}, \cite{Rom},
\cite{unpubl}:

\be \ch ^{conv}=\frac{-N}{2 \sqrt{-\mid q^{ab}\mid}} \e _{abc}\e
_{def}q^{ad}q^{be}(e^{cI}e^{fI}+\frac{1}{4}b^{cI}b^f_I) +N^a\frac{1}{2}\e
_{abc}e^{bI}b^c_I + \l ^I\cd _a e^a _I
\label{30} \ee
Now, to get exact agreement, we perform a canonical transformation in the
unified theory: $\tilde{e}^{aI}:=e^{aI}-\frac{3 i}{4 \l} b^{aI}$, and $a_{ai}$
unchanged. With this,
(\ref{20}) becomes:

\be \bar{N}\ch ^{(2)}_{YM}=\frac{-\bar{N}}{2 \sqrt{-\mid \bar{q}^{ab}\mid}}
\e _{abc}\e_{def}\bar{q}^{ad}\bar{q}^{be}(\frac{\l}{3}\tilde{e}^{cI}\tilde{e}
^f_I + \frac{3}{16\l}b^{cI}b^f_I) \label{40} \ee
So, we see that the physical Yang-Mills fields are:

\bea e^{aI}_{phys}&=&\sqrt{\frac{2 \l}{3}}\tilde{e}^{aI} \label{50} \\
 b^{aI}_{phys}&=&\sqrt{\frac{3}{2\l}}b^{aI} \nn \eea
and that the unified theory (\ref{1}) reproduces the conventional
Yang-Mills theory to lowest order.
This means that in a weak field expansion around de Sitter
spacetime, the rescaled Yang-Mills fields (\ref{50}) will be governed by the
Yang-Mills equations of motion. For a $U(1)$ Yang-Mills field, we know that
these equations of motion are Maxwell's equations which are very well
experimentally confirmed. So the question is, for what energy scales does
this unified theory predict significant corrections to the Maxwell's
equations? For a very small $\l$ (since $\l$ has dimension inverse length
square, we really mean; on length scales where $\l r^2 \ll 1$) we know that
the de Sitter metric is approximately the Minkowski metric. So, $ \bar{E}^{aA}
\bar{E}^b_A \approx \d ^{ab}$ in cartesian coordinates. The weak field
expansion is then good for

\be e^{aI}_{phys}e^b_{I,phys}\ll \frac{2 \l}{3}\d ^{ab} \label{70} \ee
With an experimental upper bound on $\l$ of $10^{-62}$ $m^{-2}$ \cite{300}
this restricts the electric field to be much weaker than $10^{-3}V/m$! This
seems to be a severe problem for this theory: it predicts large corrections
to Maxwell's equations already for rather modest field strengths. Note
however that the cosmologically constant here really just is a representative
for any slowly varying background energy density. This means that in an
experiment in a lab here on earth we must include in $\l$ all the
contributions coming from e.g thermal energy. (In room
temperate air, the heat energy-density is about $10^{-40}$ $m^{-2}$ in natural
units, which means that the restriction on the electric field increases to
$10^8 V/m$.)

Another fact that makes the importance of the value of the cosmological
constant unclear (in this context), is that in a coupling to a massive scalar
or spinor field, the mass term normally looks like a cosmological constant
term. This could mean that when e.g spinors are included, it is the spinor
mass that becomes important instead of the cosmological constant.\\ \\

Further work related to this unified theory can be found in \cite{unpubl},
where the coupling to spinors and scalar fields are studied, as well as the
static and spherically symmetric solution for gauge group $U(2)$. For related
work in (2+1)-dimensions, see \cite{2+1}.\\ \\
S.C is greatful to U.G.C (India) and Council for International Exchange of
Scholars (CIES), for giving the opportunity to visit the Center for
Gravitational Physics and Geometry, Penn State University. S.C's work was
supported in part by the grant from CIES, grant no. 17263. P.P's work was
supported by the NFR (Sweden) contract no. F-PD 10070-300, and {\it Per Erik
Lindahls
Fond, Kungliga Vetenskapsakademien}. \\ \\


\normalsize
\begin{thebibliography}{99}
\bibitem{ADM} R. Arnowitt, S. Deser and C. W. Misner, {\it Gravitation: An
introduction to current research.} (L. Witten, Ed.), (Wiley, New York, 1962).
\bibitem{Ash1}A. Ashtekar, Phys. Rev. D {\bf36} (1987) 1587.
\bibitem{ART}A. Ashtekar, J.D. Romano and R.S. Tate, Phys. Rev. D {\bf 40}
(1989) 2572.
\bibitem{thesis} P. Peld\'{a}n, "Actions for gravity, with generalizations: A
review." gr-qc/9305011, To appear in Class. Quant. Grav.
\bibitem{HKT}S. Hojman, K. Kucha\v{r} and C. Teitelboim, Ann. Phys., NY {\bf
96}
(1976) 88.
\bibitem{arbgg} P. Peld\'{a}n, Phys. Rev. D {\bf 46} (1992) R2279.
\bibitem{unpubl} S. Chakraborty and P. Peld\'{a}n, work in preparation.
\bibitem{Rom} J. D. Romano, Gen. Rel. Grav. {\bf 25} (1993) 759.
\bibitem{300} {\it General Relativity, An Einstein Centenary Survey}, (S.W.
Hawking and W. Israel, Ed.'s), (Cambridge University Press, 1979)
\bibitem{2+1} P. Peld\'{a}n, Nucl. Phys. {\bf B395} (1993) 239.
\end{thebibliography}
\end{document}